\documentclass[prd,aps,showpacs,preprintnumbers,amssymb]{revtex4}
\usepackage{graphicx}

\usepackage{epsf}

\def\e3p{$\eta \rightarrow 3 \pi$}

\begin{document}

\title{%
\hfill{\normalsize\vbox{%
\hbox{\rm SU-4252-}
 }}\\
{Chiral Nonet Mixing in
$\pi\pi$ Scattering}}

\author{Amir H. Fariborz,
$^{\it \bf a}$~\footnote[1]{Email:
 fariboa@sunyit.edu}}

\author{Renata Jora
$^{\it \bf b}$~\footnote[2]{Email:
   rjora@ifae.es          }}

\author{Joseph Schechter
 $^{\it \bf c}$~\footnote[4]{Email:
 schechte@phy.syr.edu}}

\author{M. Naeem Shahid
$^{\it \bf c}$~\footnote[5]{Email:
   mnshahid@phy.syr.edu            }}

\affiliation{$^ {\bf \it c}$ Department of Physics,
 Syracuse University, Syracuse, NY 13244-1130, USA,}
\affiliation{$^ {\bf \it a}$ Department of 
Engineering, Sciences and Mathematics, 
State University of New York, Institute of Technology,Utica, NY13504-3050, USA,}
\affiliation{$^ {\bf \it b}$ Grup de Fisica Teorica and IFAE,
  Universitat Autonoma de Barcelona, E-08193 Bellaterra (Barcelona), Spain,}
\date{\today}

\begin{abstract}
Pion pion scattering is studied in a generalized
linear sigma model  which contains two scalar nonets
(one of quark-antiquark type and the other of
diquark-antidiquark type)
and two corresponding pseudoscalar nonets.
An interesting feature concerns the mixing of the
four isosinglet scalar mesons which yield poles in
the scattering amplitude.
Some realism is introduced by enforcing exact unitarity
via the K-matrix method.
 It is shown that a reasonable
agreement with experimental data is obtained up to
about 1 GeV.  The poles in the unitarized
scattering amplitude are studied in some detail.
The lowest pole clearly represents the sigma meson (or
$f_0(600)$) with a mass and decay width around 500
MeV. The second pole
invites comparison with the $f_0(980)$ which has
a mass around 1 GeV and decay width around 100
MeV. The third and fourth poles, resemble
some of the isosinglet state in the complicated
1-2 GeV region.
Some comparison is made to the situation in the
usual SU(3) linear sigma model with a single scalar
nonet.
 \end{abstract}


\maketitle
\section{introduction}

Although the exact nature of the low lying scalar mesons has
been a topic of intense debate,  the fact
that these states play important roles
in our understanding of low-energy QCD seems to
be shared by all.
 Various
models have been put forward for the properties
of the scalar mesons. A general discussion of the experimental
situation on light scalars is given in ref. \cite{pdg}. Some characteristic treatments of the last twenty years are given in refs. \cite{vanBev}-\cite{fjss11}.
In particular, an important
four-quark (i.e. two quarks and two antiquarks) component, as was first proposed in the
MIT bag model \cite{j77},  seems to explain
some
of their unusual properties such as the reversed mass spectrum.
It has also been pointed out \cite{BFS3}, \cite{BFMNS01}, \cite{mixing},
\cite{tsm}, \cite{thermo}, \cite{FJS05}-\cite{fjs09}, that four
quark components alone are not sufficient for
understanding the physical parameters
of these states but seems to rquire a scenario based on an underlying
mixing between quark-antiquark nonets and
nonets containing two quarks as well as two anti quarks.
 A simple picture for scalar states
below 2 GeV then seems to emerge. Amusingly, this mixing
 \cite{fjs09} automatically
leads to light scalars that are dominantly of  two quark- two antiquark
 nature and light
conventional pseudoscalars that are, as expected
from established phenomenology,  dominantly
of quark-antiquark nature.

As some of the light scalar mesons (such as the
$\sigma$ and the $\kappa$) are very broad their
Lagrangian masses (``bare'' masses) are considerably
different from their pole locations in appropriate
scattering amplitudes.
The work of \cite{fjs09} investigated in
detail the ``bare'' mass spectrum of these
states, as well as their internal structure by
fitting the predictions of the model for
various low-energy parameters to known
experimental data.  This fixed all Lagrangian
parameters (to the leading order).  In this
paper, we work with the same
Lagrangian (i.e. without introducing any new
parameters) and investigate the effect of
unitarization on the isosinglet, zero angular momentum
partial wave pi-pi scattering amplitude computed at tree order.

  We will treat the
 pion scattering
amplitude unitarization by using the
K-matrix method.     As the model
involves two nonets of scalars,
there are altogether four isosinglet scalar
mesons (two from each nonet) that contribute as
poles in the pion scattering amplitude.
Therefore the K-matrix unitarization
has to deal with all four poles at
the same time resulting in a more involved
version of the conventional single-pole
K-matrix unitarization.

     The advantages of the K-matrix approach to unitarization
     are that it does not introduce any new parameters and that
     it forces exact unitarity. It is plausible since if one starts
     from a pure pole in the partial wave amplitude, one ends up with a
     pure Breit Wigner shape. A disadvantage is that it neglects, in the
     simple version we use, the effects of the opening of thresholds like
     the $K{\bar K}$ on the pi pi amplitude. This is not
expected to be too serious for our initial appraisal here.

In Sec. II we give a brief review of the
Lagrangian and the relevant formulas, followed
by the K-matrix unitarization of the scattering
amplitude in Sec. III and a summary and
discussion of the results in Sec. IV.


\section{Brief Review of the model}


   The model employs the 3$\times$3 matrix
chiral nonet fields:
\begin{equation}
M = S +i\phi, \hskip 2cm
M^\prime = S^\prime +i\phi^\prime.
\label{sandphi}
\end{equation}
The matrices $M$ and $M'$ transform in the same way under
chiral SU(3) transformations but
may be distinguished by their different U(1)$_A$
transformation properties. $M$ describes the ``bare"
 quark antiquark scalar and pseudoscalar nonet fields while
$M'$ describes ``bare" scalar and pseudoscalar fields
containing two quarks and two antiquarks. At the
symmetry level in which we are working,
it is unnecessary to
further specify the four quark field configuration.
The four quark field may, most generally,
 be imagined as some linear
combination of a diquark-antidiquark  and a
``molecule" made of two quark-antiquark ``atoms".

The general Lagrangian density which defines our model is
\begin{equation}
{\cal L} = - \frac{1}{2} {\rm Tr}
\left( \partial_\mu M \partial_\mu M^\dagger
\right) - \frac{1}{2} {\rm Tr}
\left( \partial_\mu M^\prime \partial_\mu M^{\prime \dagger} \right)
- V_0 \left( M, M^\prime \right) - V_{SB},
\label{mixingLsMLag}
\end{equation}
where $V_0(M,M^\prime) $ stands for a function made
from SU(3)$_{\rm L} \times$ SU(3)$_{\rm R}$
(but not necessarily U(1)$_{\rm A}$) invariants
formed out of
$M$ and $M^\prime$.

 As we previously discussed \cite{1FJS07}, the
 leading choice of terms
corresponding
to eight or fewer underlying quark plus antiquark lines
 at each effective vertex
reads:
\begin{eqnarray}
V_0 =&-&c_2 \, {\rm Tr} (MM^{\dagger}) +
c_4^a \, {\rm Tr} (MM^{\dagger}MM^{\dagger})
\nonumber \\
&+& d_2 \,
{\rm Tr} (M^{\prime}M^{\prime\dagger})
     + e_3^a(\epsilon_{abc}\epsilon^{def}M^a_dM^b_eM'^c_f + h.c.)
\nonumber \\
     &+&  c_3\left[ \gamma_1 {\rm ln} (\frac{{\rm det} M}{{\rm det}
M^{\dagger}})
+(1-\gamma_1){\rm ln}\frac{{\rm Tr}(MM'^\dagger)}{{\rm
Tr}(M'M^\dagger)}\right]^2.
\label{SpecLag}
\end{eqnarray}
     All the terms except the last two (which mock up the axial anomaly)
      have been chosen to also
possess the  U(1)$_{\rm A}$
invariance.
The symmetry breaking term which models the QCD mass term
takes the form:
\begin{equation}
V_{SB} = - 2\, {\rm Tr} (A\, S)
\label{vsb}
\end{equation}
where $A=diag(A_1,A_2,A_3)$ are proportional to
  the three light quark
masses.
The model allows for two-quark condensates,
$\alpha_a=\langle S_a^a \rangle$ as well as
four-quark condensates
$\beta_a=\langle {S'}_a^a \rangle$.
Here we assume \cite{SU} isotopic spin
symmetry so A$_1$ =A$_2$ and:
\begin{equation}
\alpha_1 = \alpha_2  \ne \alpha_3, \hskip 2cm
\beta_1 = \beta_2  \ne \beta_3
\label{ispinvac}
\end{equation}

 We also need the ``minimum" conditions,
\begin{equation}
\left< \frac{\partial V_0}{\partial S}\right> + \left< \frac{\partial
V_{SB}}{\partial
S}\right>=0,
\quad \quad \left< \frac{\partial V_0}{\partial S'}\right>
=0.
\label{mincond}
\end{equation}

There are twelve parameters describing the Lagrangian and the
vacuum. These include the six coupling constants
 given in Eq.(\ref{SpecLag}), the two quark mass parameters,
($A_1=A_2,A_3$) and the four vacuum parameters ($\alpha_1
=\alpha_2,\alpha_3,\beta_1=\beta_2,\beta_3$).The four minimum
equations reduce the number of needed input parameters to
eight.

Five of these eight are supplied by the following
masses together with the pion decay constant:
\begin{eqnarray}
 m[a_0(980)] &=& 984.7 \pm 1.2\, {\rm MeV}
\nonumber
\\ m[a_0(1450)] &=& 1474 \pm 19\, {\rm MeV}
\nonumber \\
 m[\pi(1300)] &=& 1300 \pm 100\, {\rm MeV}
\nonumber \\
 m_\pi &=& 137 \, {\rm MeV}
\nonumber \\
F_\pi &=& 131 \, {\rm MeV}
\label{inputs1}
\end{eqnarray}
Because $m[\pi(1300)]$ has such a large uncertainty,
we will, as previously, examine predictions
depending on the choice of this mass
within its experimental range.
The sixth input will be taken as the light
``quark mass ratio" $A_3/A_1$, which will
be varied over an appropriate range.
 The remaining two inputs will be taken from the
 masses of the four (mixing) isoscalar, pseudoscalar
mesons. This mixing is characterized by a 4
$\times$ 4 matrix
$M_\eta^2$. A practically convenient choice is to consider
Tr$M_\eta^2$ and det$M_\eta^2$ as the inputs.

Given these inputs there are a very large number of
predictions. At the level of the quadratic terms in the
Lagrangian, we predict all the remaining masses
 and decay constants as well
as the angles describing the mixing between each of
($\pi,\pi'$),
($K,K'$), ($a_0,a_0'$), ($\kappa,\kappa'$) multiplets
and each of the 4$\times$4
isosinglet mixing matrices
 (each formally described by six angles).

In the case of the I=0 scalars there are four
 particles which mix with each other; the squared
mass matrix then takes the form:

\begin{equation}
\left(  X_0^2  \right) =
 \left[
  \begin{array}{cccc}
4  \,   e_3^a  \,  {\it \beta_3}  -  2  \,
c_2  +  12  \,   c_4^a  \,   \alpha_1^2
&
4  \, \sqrt{2}\,  e_3^a  \,  \beta_1
&
4  \,  e_3^a  \,   \alpha_3
&
4 \, \sqrt{2}\,    e_3^a  \,  \alpha_1
\\
4  \, \sqrt{2}\,  e_3^a   \,
\beta_1
&
-2  \,  c_2  + 12  \,  c_4^a  \, \alpha_3^2
&
4  \,  \sqrt{2}\, e_3^a  \,
\alpha_1
&
0
\\
4  \,  e_3^a  \,   \alpha_3
&
4  \,  \sqrt{2}\,  e_3^a   \,
\alpha_1
&
2  \,   d_2
&
0
\\
4  \,  \sqrt{2}\, e_3^a  \,
\alpha_1
&
0
&
0
&
2 \, d_2
\end {array}
\right]
\end{equation}

For this matrix the basis states are consecutively,

\begin{eqnarray}
f_a&=&\frac{S^1_1+S^2_2}{\sqrt{2}} \hskip .7cm
n{\bar n},
\nonumber  \\
f_b&=&S^3_3 \hskip .7cm s{\bar s},
\nonumber    \\
f_c&=&  \frac{S'^1_1+S'^2_2}{\sqrt{2}}
\hskip .7cm ns{\bar n}{\bar s},
\nonumber   \\
f_d&=& S'^3_3
\hskip .7cm nn{\bar n}{\bar n}.
\label{fourbasis}
\end{eqnarray}
    The non-strange (n) and strange (s) quark content
for each basis state has been listed at the end of
each line above.


\section{Pion Scattering Amplitude}

     Some initial discussion of the pion scattering in this model
     was given in refs. \cite{1FJS07} and
     \cite{2FJS07}.
The tree level $\pi\pi$ scattering amplitude is:
\begin{equation}
A(s,t,u) = - {g\over 2}
+ \sum_i {{g_i^2}\over {m_i^2 - s}},
\label{Astu}
\end{equation}
where the four point coupling constant is related to the ``bare''
four-point couplings as:
\begin{equation}
g
 =
\left\langle
{{\partial^4 V}
\over
{\partial \pi^+ \, \partial \pi^- \, \partial \pi^+ \,
\partial \pi^-}}
\right\rangle
 =
\sum_{A,B,C,D}
\left\langle
{{\partial^4 V}
\over
{
 \partial (\phi_1^2)_A \,
 \partial (\phi_2^1)_B \,
 \partial (\phi_1^2)_C \,
 \partial (\phi_2^1)_D
}}
\right\rangle \,
(R_\pi)_{A1} \,
(R_\pi)_{B1} \,
(R_\pi)_{C1} \,
(R_\pi)_{D1}
\end{equation}
where the sum is over ``bare'' pions and
$A, B, \cdots$ = 1, 2 with 1 denoting nonet $M$ and 2
denoting nonet $M'$.    $R_\pi$ is the pion
rotation matrix (given, for typical parameters in \cite{fjs09}.

The physical scalar-psedudoscalar-pseudoscalar
couplings are related to the bare couplings:
\begin{equation}
g_i =
\left\langle
{{\partial^3 V}
\over
{\partial f_i \, \partial \pi^+ \, \partial \pi^-}}
\right\rangle
=
\sum_{M,A,B}
\left\langle
{{\partial^3 V}
\over
{
 \partial f_M
 \partial (\phi_1^2)_A \,
 \partial (\phi_2^1)_B
}}
\right\rangle
(L_0)_{M i} \,
(R_\pi)_{A1} \,
(R_\pi)_{B1}
\end{equation}
where $A$ and $B$ = 1,2 and $M$=1,2,3 and 4
and respectively represent the four bases in Eq.
(\ref{fourbasis}). $L_0$ is the isosinglet scalar
rotation matrix.

The only non-vanishing ``bare'' four-point and
three-point couplings
are:
\begin{eqnarray}
\left\langle
{{\partial^4 V}
\over
{
 \partial (\phi_1^2)_1 \,
 \partial (\phi_2^1)_1 \,
 \partial (\phi_1^2)_1 \,
 \partial (\phi_2^1)_1
}}
\right\rangle
&=& 8\, c_4^a\\
\left\langle
{{\partial^3 V}
\over
{
 \partial f_a\,
 \partial (\phi_1^2)_1 \,
 \partial (\phi_2^1)_1
}}
\right\rangle
&=& 4\, \sqrt{2}\, c_4^a \, \alpha_1 \nonumber \\
\left\langle
{{\partial^3 V}
\over
{
 \partial f_b\,
 \partial (\phi_1^2)_1 \,
 \partial (\phi_2^1)_2
}}
\right\rangle
&=&
\left\langle
{{\partial^3 V}
\over
{
 \partial f_b\,
 \partial (\phi_1^2)_2 \,
 \partial (\phi_2^1)_1
}}
\right\rangle
=
\left\langle
{{\partial^3 V}
\over
{
 \partial f_c\,
 \partial (\phi_1^2)_1 \,
 \partial (\phi_2^1)_1
}}
\right\rangle
= 4\, e_3^a
\end{eqnarray}

Now we project Eq. (\ref{Astu}) to the I=J=0 partial wave
amplitude.
The K-matrix unitarization of this "Born"  scattering
amplitude ${T_0^0}^B$ defines the unitary partial wave amplitude
\begin{equation}
T_0^0 = {  {T_0^0}^B \over { 1 - i\, {T_0^0}^B} }
\label{T00_unitary}
\end{equation}
wherein:
\begin{equation}
{T_0^0}^B = T_\alpha + \sum_i  {  {T_\beta^i } \over
{m_i^2 - s}}
\label{T00B}
\end{equation}
with:
\begin{equation}
T_\alpha =
{1\over 64 \pi}
\sqrt{1 - {4 m_\pi^2\over s}}\,
\left[-5 g_4 +
  { 1 \over {p_\pi^2}}\,
   \sum_i g_i^2\,  {\rm ln} \left(1 +  {{4
p_\pi^2}\over m_i^2} \right)
\right]
\end{equation}

\begin{equation}
T_\beta^i =
{3\over 32 \pi}
\sqrt{1 - {4 m_\pi^2\over s}}\, g_i^2
\end{equation}

\begin{equation}
p_\pi =
{1\over 2} \sqrt{s - 4 m_\pi^2}
\end{equation}

\section{Comparison with experiment}

\begin{figure}
\begin{center}
\vskip 1cm
\epsfxsize = 12cm
 \epsfbox{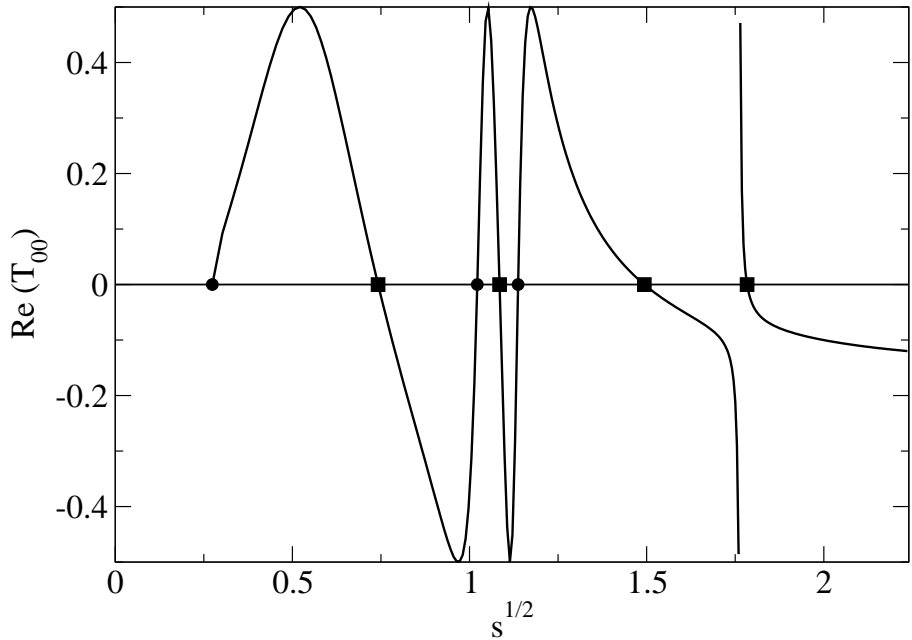}
\end{center}
\caption[]{%
 The real
part of the unitarized $\pi\pi$ scattering amplitude for a typical choice of parameters.  The four squares
 correspond to the poles in $T_0^{0B}$.The circles correspond to
 locations where $T_0^{0B}$ = 0.
}
\label{F_generic_T00}
\end{figure}

 For comparison with experiment it is convenient to focus on the
 real part of the partial wave scattering amplitude in Eq. (\ref{T00_unitary}).
 For typical values of parameters we find the behavior illustrated in Fig. (\ref{F_generic_T00}).
The zeros which occur can be understood as follows. First, they can result from
 a zero of $T_0^{0B}$. Such a zero occurs at threshold,
for example. Secondly, a zero can also result from the poles in
$T_0^{0B}$ at $s=m_i^2$ in Eq.(\ref{T00B}) corresponding to the ``bare" masses.

\begin{figure}
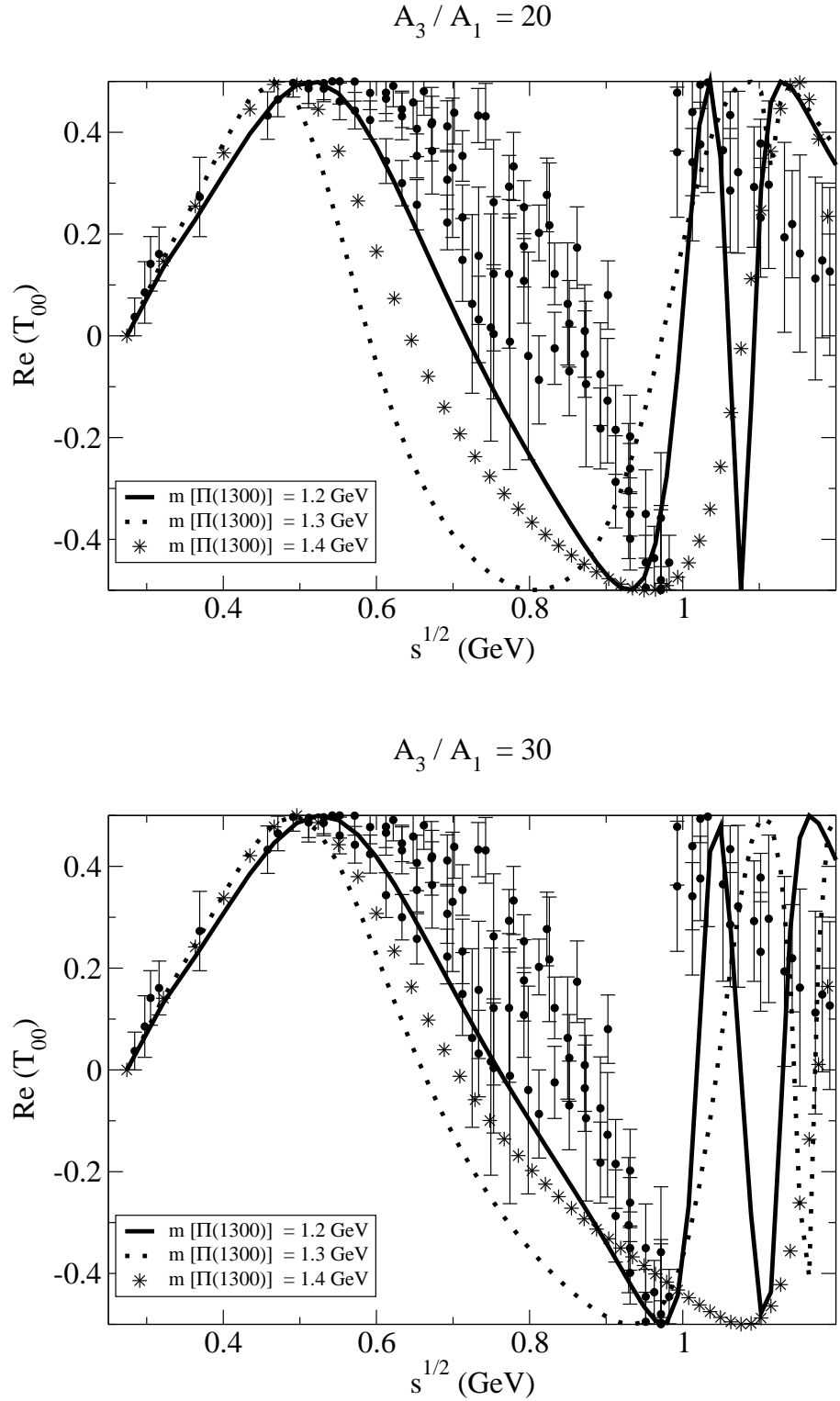

\begin{center}
\vskip 1cm
\epsfxsize = 12cm
 \epsfbox{fig2a.eps}
\vskip 1cm

\epsfxsize = 12cm
 \epsfbox{fig2b.eps}
\end{center}
\caption[]{%
Real part of unitarized scattering amplitude
for two values of $A_3/A_1$ and three choices
of $m[\Pi(1300)]$. }
\label{F_ReT00_unitary}
\end{figure}

 We
compare the predictions of our model for the
scattering amplitude with the corresponding
experimental data up to about 1.2 GeV in Fig \ref{F_ReT00_unitary}
for two values of the SU(3) symmetry breaking
parameter $A_3/A_1$
and three choices of the only roughly
known "heavy pion" mass $m[\Pi(1300)]$.
One sees that, without using any new parameters,
the mixing mechanism of \cite{fjs09}
predicts the scattering amplitude in reasonable
 qualitative agreement with the experimental
data up to
around 1 GeV. This provides some  support for the
validity of this mixing mechanism.

 For interpretation of the physical resonances it is
 conventional to look at the pole positions in the complex plane of the
 analytically continued expression for $T_0^0$.
We examine these physical pole positions by
solving for the complex roots of the
denominator of the  K-matrix unitarized amplitude
Eq. (\ref{T00_unitary}):
\begin{equation}
{\cal D} (s) = 1 - i\, T_0^{0B} = 0
\label{pole_eq}
\end{equation}
with $T_0^{0B}$ given by Eq. (\ref{T00B}).   We
search for solutions, $s^{(j)}
= s_r^{(j)}
+ i s_i^{(j)} = m_j^2 - i m_j \Gamma_j$ of this equation, where
$m_j$ and $\Gamma_j$ are interpreted as the
mass and decay width of the j-th physical
resonance ( which would hold for small $\Gamma$ ). A
first natural attempt would be to simultaneously solve the two equations:
\begin{eqnarray}
{\rm Re}{\cal D} \left(s_r, s_i\right) = 0 \nonumber \\
{\rm Im}{\cal D} \left(s_r, s_i\right) = 0
\label{ReandIm}
\end{eqnarray}
However, this approach turns out to be rather complicated to be
implement. A more efficient numerical approach is to consider the single equation
involving only positive quantities:
\begin{equation}
{\cal F} \left(s_r, s_i\right) =
\left| {\rm Re}
\left(
{\cal D} (s_r, s_i)
\right)\right| +
\left| {\rm Im}
\left(
{\cal D} (s_r, s_i)
\right)\right|=0
\label{F_srsi}
\end{equation}
A search of parameter space leads to four solutions for the
pole positions \cite{foot}. As an example, for the choice of $A_3/A_1$ = 30
and $m[\Pi(1300)]=$ 1.215 GeV, the function ${\cal F}$
is plotted over the complex plane around the
first pole. We see a clear local minimum
at which the function is zero, hence pointing
to a solution of Eq. (\ref{pole_eq}). Similarly, other
areas of the complex plane are searched and
altogether four poles are found. The results
are given in Table \ref{T_poles} for
$m[\Pi(1300)]=$ 1.215 GeV and two choices of
$A_3/A_1$ = 20 and 30.  For each choice we see
that this model predicts a light and broad
scalar meson below 1 GeV which is a clear
indication of $f_0(600)$ or $\sigma$.
We see that the characteristics of the
second predicted state around 1
GeV are close to those expected for $f_0(980)$.   The third and the fourth
predicted states should correspond to two of
$f_0(1370)$, $f_0(1500)$ and $f_0(1710)$.

\begin{figure}
\begin{center}
\vskip 1cm
\epsfxsize = 12cm
 \epsfbox{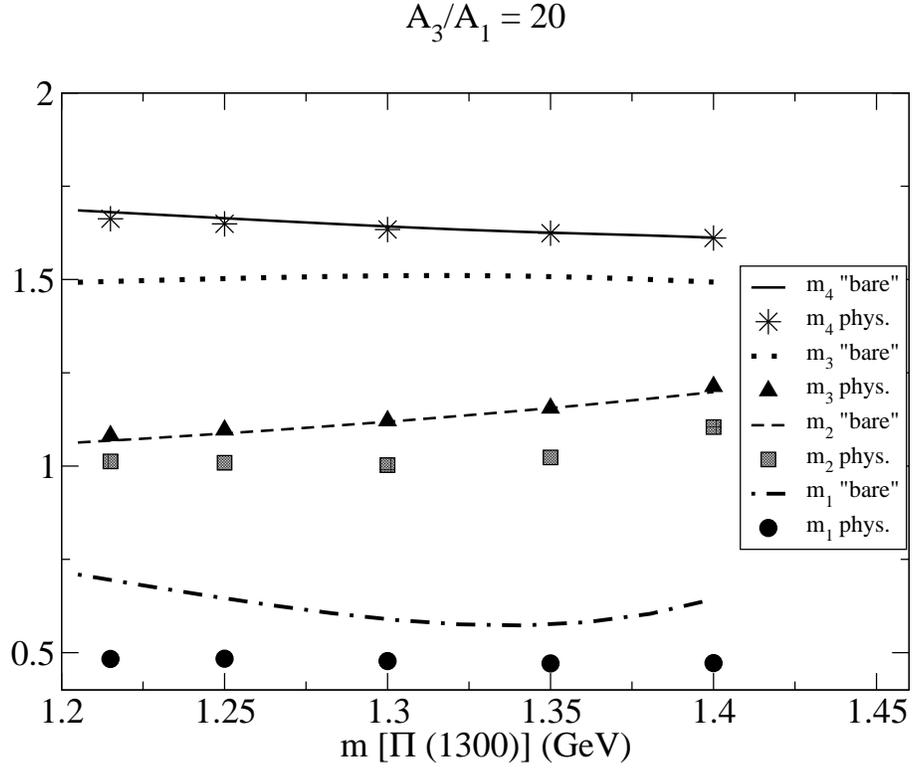}
\vskip 1cm

\epsfxsize = 12cm
 \epsfbox{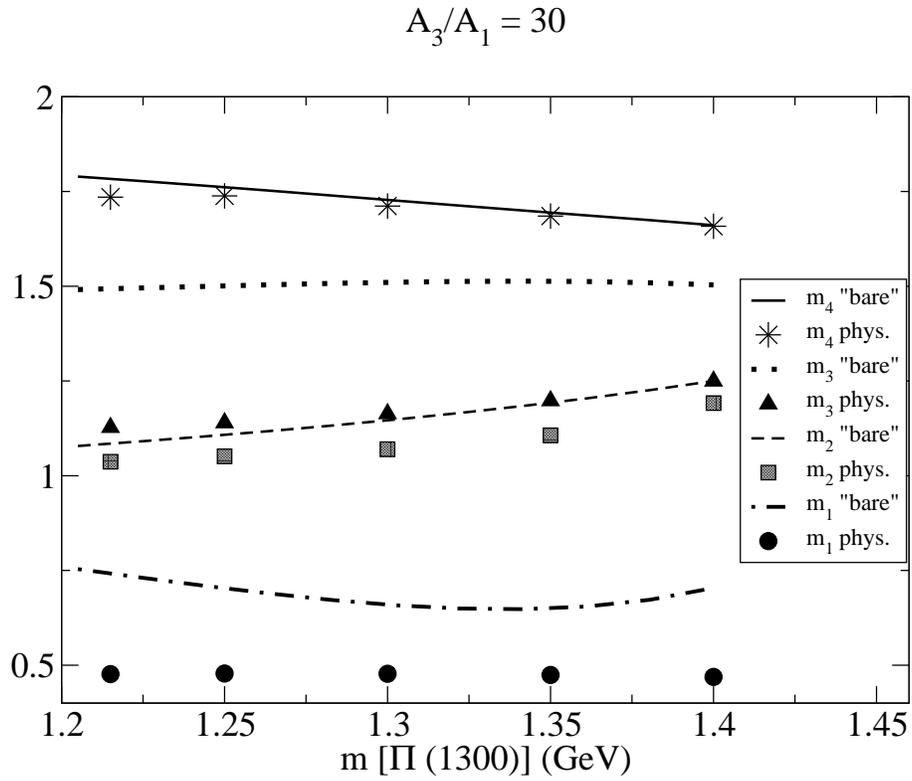}
\end{center}
\caption[]{%
Predicted physical masses are compared with the
``bare'' masses for two values of $A_3/A_1$
over the experimental range of $m[\Pi(1300)]$. }
\label{F_m_vs_mpip}
\end{figure}

\begin{figure}
\begin{center}
\vskip 1cm
\epsfxsize = 12cm
 \epsfbox{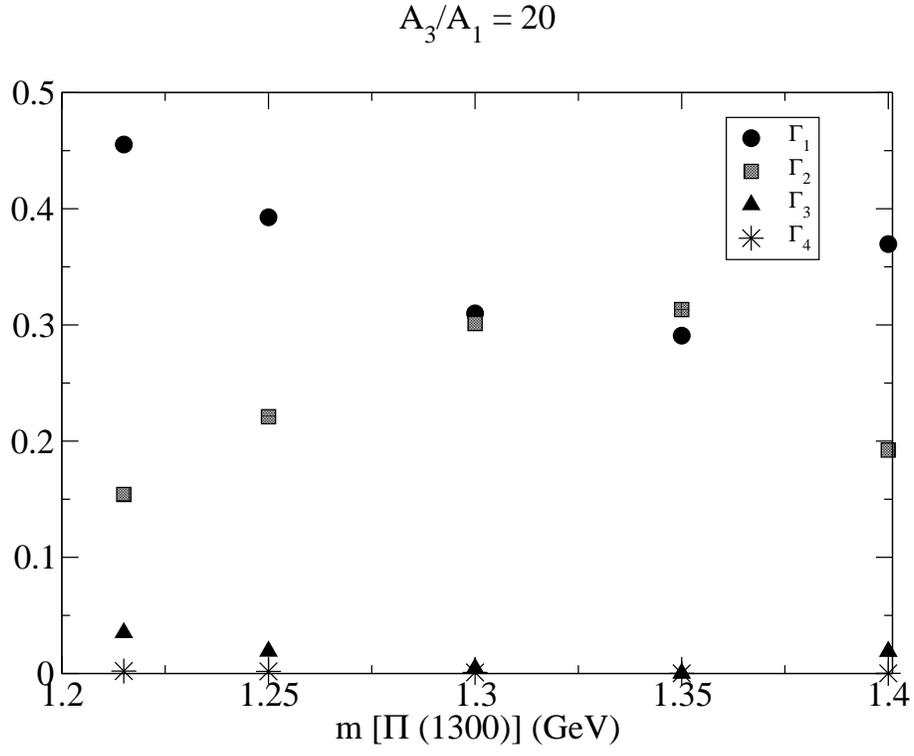}
\vskip 1cm

\epsfxsize = 12cm
 \epsfbox{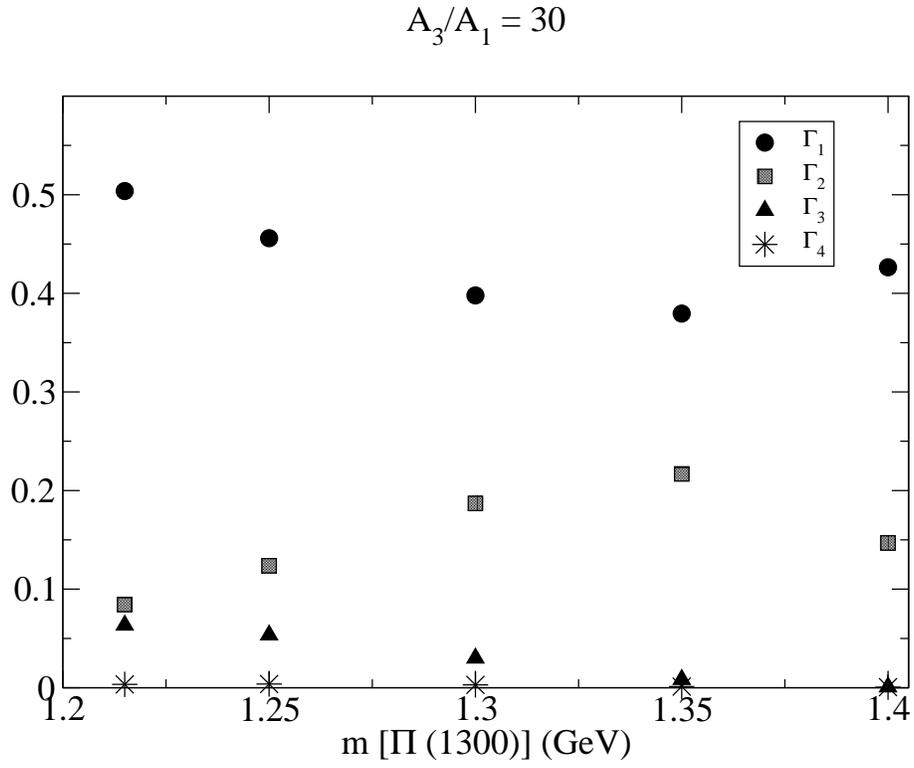}
\end{center}
\caption[]{%
Predicted decay widths for two values of
$A_3/A_1$ over the experimental range of
$m[\Pi(1300)]$.}
\label{F_G_vs_mpip}
\end{figure}

We have performed the same analysis over the
range of the parameter $m[\Pi(1300)]$ = 1.2 - 1.4 GeV,
and for two choices of $A_3/A_1$ = 20 and 30.
The physical masses and the decay widths are
given in Figs. \ref{F_m_vs_mpip} and
\ref{F_G_vs_mpip}, respectively.   The effect
of the unitarization can be seen in Fig.
\ref{F_m_vs_mpip} where the physical masses are
compared with the ``bare'' masses; the
unitarization reduces the mass, particularly
for the first and the third predicted states.

\begin{figure}
\begin{center}
\vskip 1cm
\epsfxsize = 6cm
\epsfbox{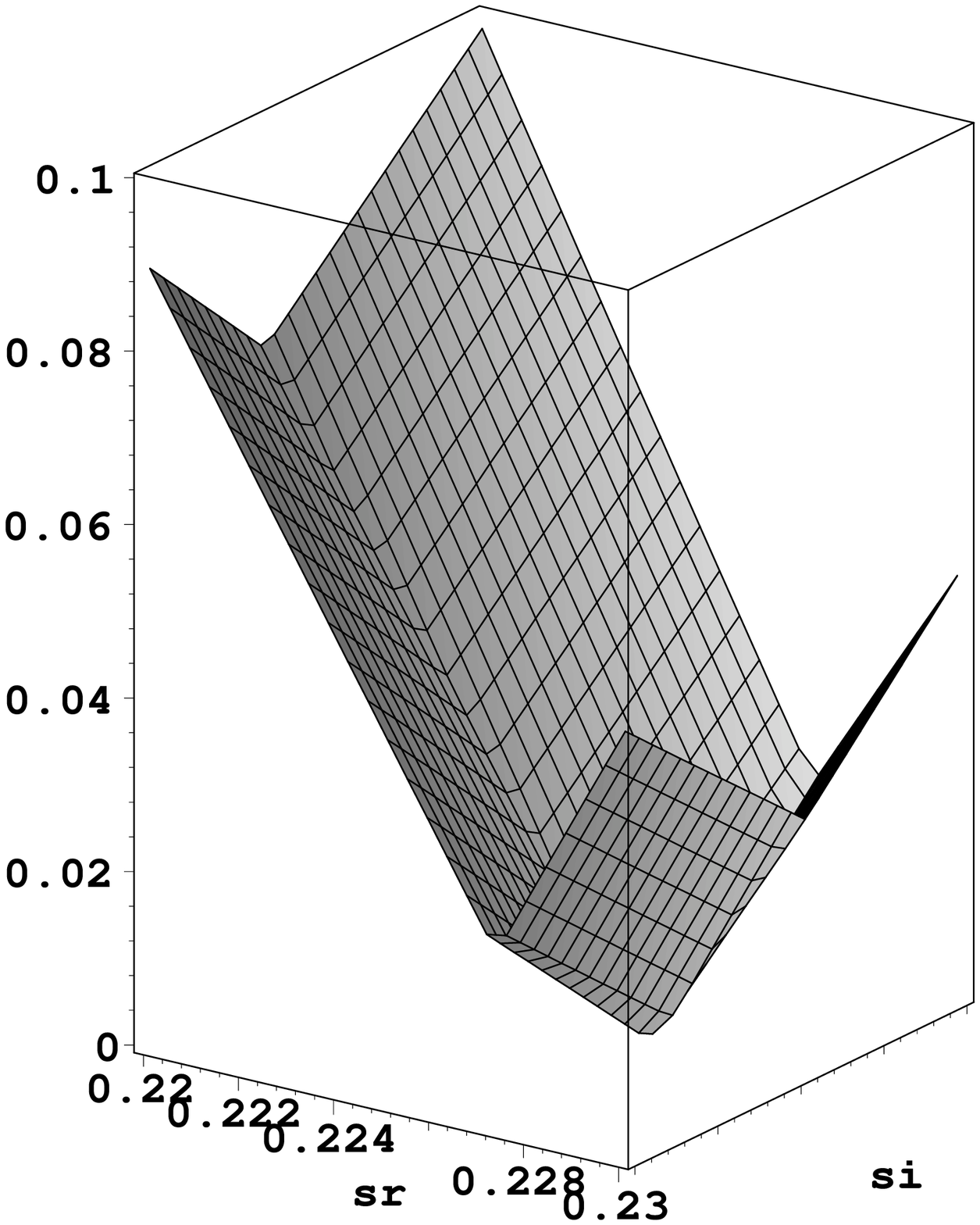}
\hskip 1cm
\epsfxsize = 6cm
\epsfbox{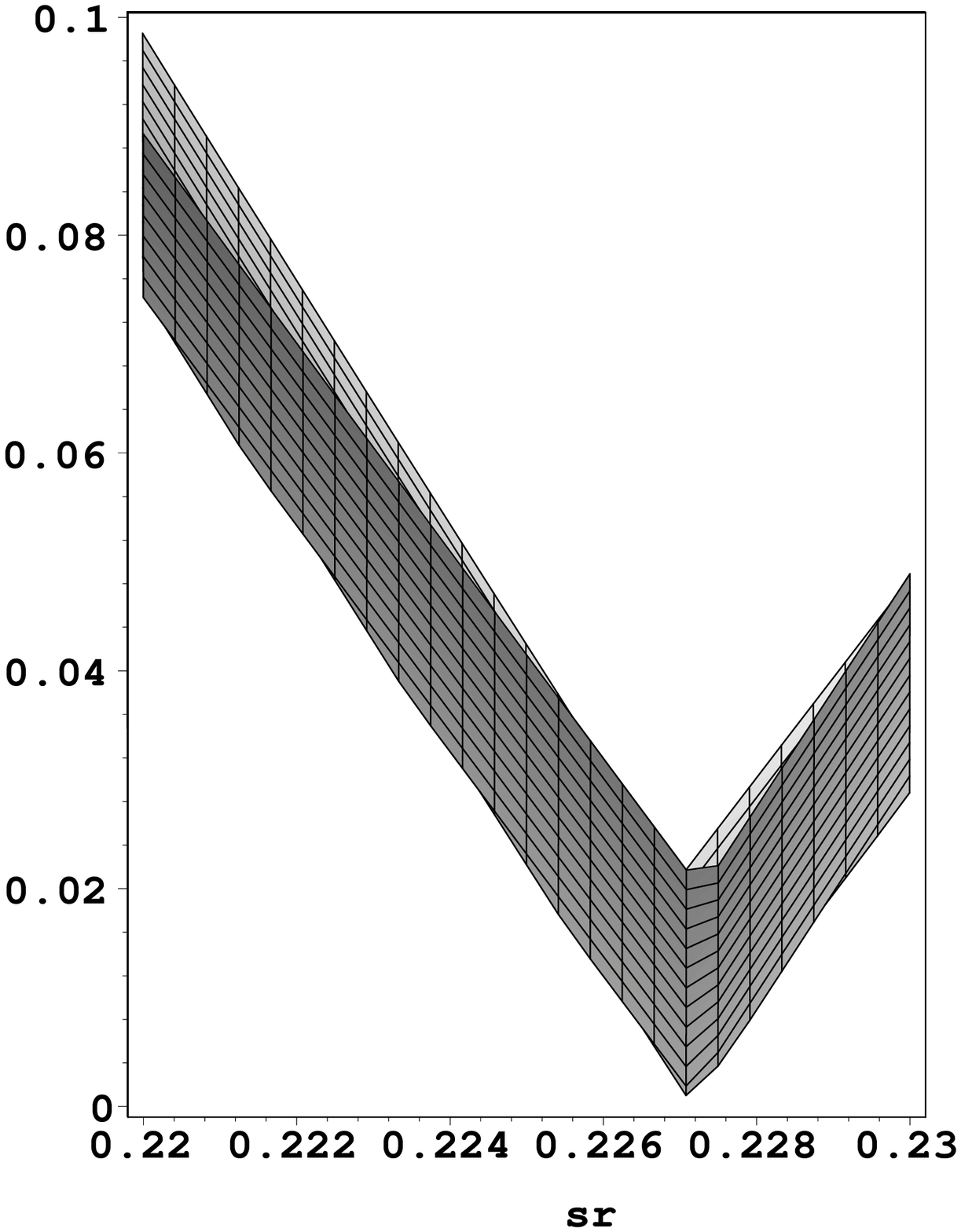}
\vskip 2cm
\epsfxsize = 6cm
\epsfbox{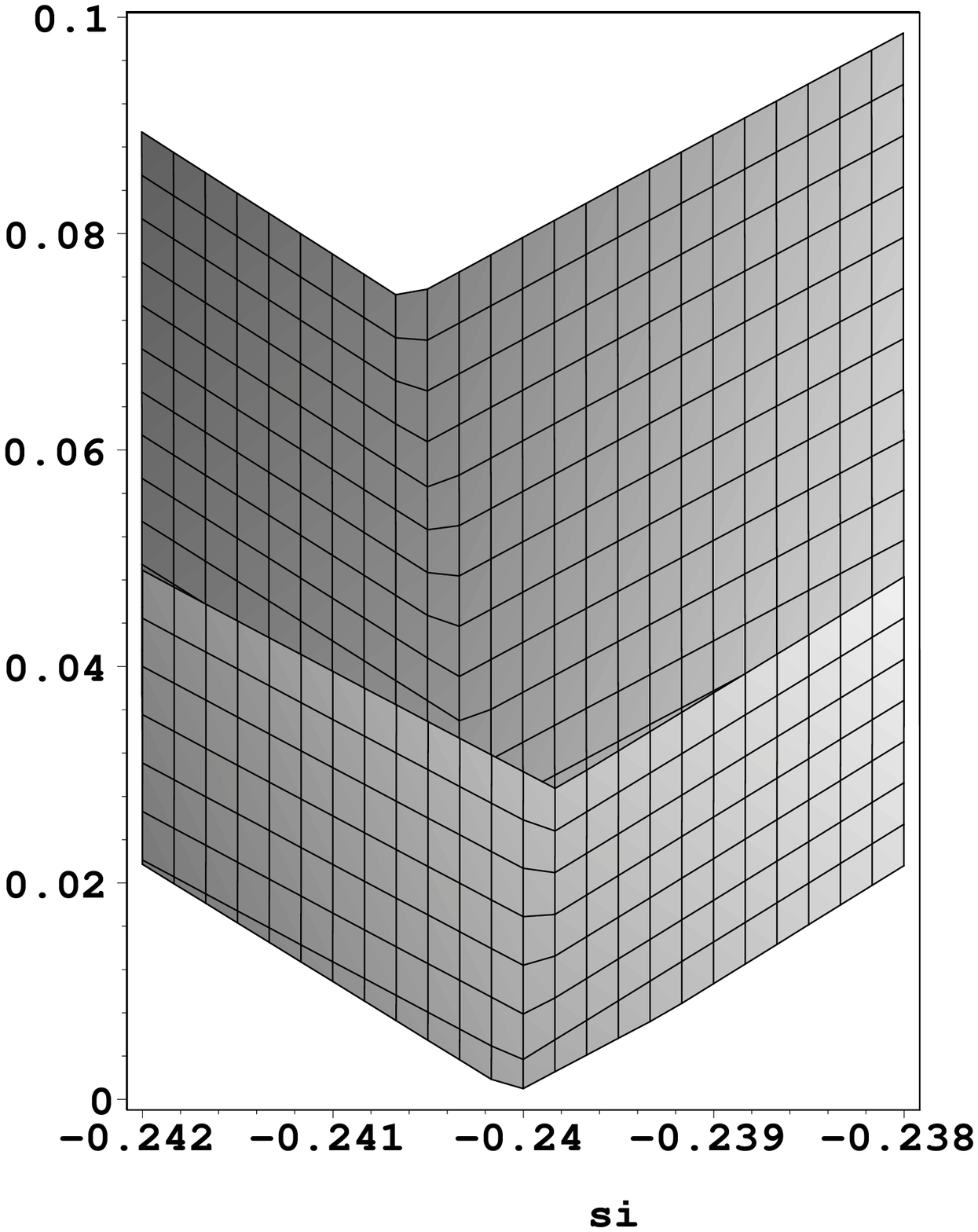}

\end{center}
\caption[]{
The local minimum of function ${\cal F}(s_R,s_I)$
defined in Eq. (\ref{F_srsi}) at the position
of the lightest isosinglet
scalar pole in the
complex $s$ plane for $m[\Pi(1300)]=1.215$ GeV
and $A_3/A_1$ = 30.
Top left is the plot of function ${\cal
F}(s_R,s_I)$ vs $s_R$ and $s_I$, followed by
projection of this function onto ${\cal
F}$-$s_R$
and onto ${\cal F}$-$s_I$ planes.
}
 \label{F_poles}
\end{figure}

\begin{table}[htbp]
\begin{center}
\begin{tabular}{c||c|c||c|c}
\hline \hline
Pole & Mass (MeV) & Width (MeV) & Mass (MeV) &
Width (MeV)
\\ \hline
1 & 483 & 455 & 477 & 504
\\ \hline
2 & 1012& 154 & 1037 & 84
\\ \hline
3 & 1082 & 35 & 1127 & 64
\\ \hline
4 & 1663 & 2.1 & 1735 & 3.5
\\ \hline
\hline
\end{tabular}
\end{center}
\caption[]{The physical mass and decay width of
the isosinglet scalar states, with $m[\Pi(1300)]
=1.215$ GeV and with $A_3/A_1$ = 20 (the
first two columns) and with $A_3/A_1$ = 30
(the last two columns).}
\label{T_poles}
\end{table}

 \section{Concluding discussion}
 
       We studied the predictions of the real (mass) and imaginary (width) parts of the pion scattering 
       amplitude poles representing  
  the isoscalar scalar singlets in a chiral model
       containing not only the usual pseudoscalar and scalar nonets describing quark-
       antiquark bound states but also pseudoscalar and scalar nonets describing states 
       with the same quantum numbers but constructed out of two quarks and two antiquarks in 
       a general way. The physical particles correspond to mixtures of these two types.

     In this model there are four scalars so the process is technically complicated.
      No new parameters
     were introduced here, either for the model itself or to treat
      the scattering. The model has been studied in a number of previous papers with
      the parameter determination
      culminating in ref. \cite{fjs09}.
      
      The fact that the comparison with the experimental scalar candidates, as discussed in 
      the last section, is reasonable is in itself a non trivial conclusion.
      A numerical technique to facilitate this result was presented in the last section. Also the fact
      that the simple single channel K-matrix unitarization (using no new parameters) seems to work may be
      useful to point out. Presumably the results would be improved if the effect of the $K$ - $\bar{K}$
      channel were to be included.  Mixing with a possible glueball state is another relevant effect.
      
      The worst prediction seems to be the too low mass value for pole 3. We note
       from Fig. \ref{F_m_vs_mpip} that there is a relatively large difference between the 
       "bare" mass and the pole mass in this case. The inclusion of the $K$ - $\bar{K}$ threshold 
       effects may improve this feature.
       
       It may also be interesting to compare the predictions of pole 1 and pole 2 with those calculated
       in a similar manner using the single M SU(3) sigma model \cite{BFMNS01}. The agreement is quite 
       good. However, in that model, the result was calculated using the most general form of the 
       interaction potential involving the field matrix M; an attempt to just use the ``renormalizable" terms
       did not give as good a result. In the present case it was not necessary to introduce any additional terms 
       in the Lagrangian to get good results for the pi pi scattering.

\section*{Acknowledgments} 
\vskip -.5cm 
We are
happy to thank A. Abdel-Rehim, D. Black, M.
Harada, S. Moussa, S. Nasri and F. Sannino for
many helpful related discussions. The work of
A. H. Fariborz has been partially supported by
the NSF Grant 0854863 and by a 2011 grant from 
the Office of the Provost, SUNYIT. The work of 
R.Jora has
been supported by CICYT-FEDEF-FPA 2008-01430.
The work of
 J.Schechter and M. N. Shahid was supported in part by the U.
S. DOE under Contract no. DE-FG-02-85ER 40231.


\begin{thebibliography}{10}
\bibitem{foot} We have double checked the results by developing an approximate analytical approach
in which the amplitude is unitarized locally in the neighbourhood of each resonance.
\end{thebibliography}
  \end{document}